%% file: sample-base.tex
\documentclass[conference]{IEEEtran}
\IEEEoverridecommandlockouts
\usepackage{cite}

\usepackage{csquotes}
\usepackage{url}
\usepackage{amsmath,amssymb,amsfonts}
\usepackage{algorithmic}
\usepackage{graphicx}
\usepackage{textcomp}
 \usepackage[T1]{fontenc}
  \usepackage[nomath]{lmodern}

\usepackage{xcolor}
\AtBeginDocument{%
  \providecommand\BibTeX{{%
    \normalfont B\kern-0.5em{\scshape i\kern-0.25em b}\kern-0.8em\TeX}}}

\def\BibTeX{{\rm B\kern-.05em{\sc i\kern-.025em b}\kern-.08em
    T\kern-.1667em\lower.7ex\hbox{E}\kern-.125emX}}
\usepackage{fancyhdr}
\usepackage{kantlipsum}
\fancypagestyle{plain}{
\fancyhf{}
\fancyhead[C]{Conference on \LaTeX} 

}
\usepackage{eso-pic}
\begin{document}
\AddToShipoutPictureBG*{
\AtPageUpperLeft{
\setlength\unitlength{1in}
\hspace*{\dimexpr0.5\paperwidth\relax}
\makebox(0,-0.75)[c]{\textbf{2022 IEEE/ACM International Conference on Advances in Social
Networks Analysis and Mining (ASONAM)}}}}

\title{Is Twitter Enough? Investigating Situational Awareness in Social and Print Media during the Second COVID-19 Wave in India
\\
}
\author{\IEEEauthorblockN{Ishita Vohra, Meher Shashwat Nigam, Aryan Sakaria, Amey Kudari, Dr. Nimmi Rangaswamy}
\IEEEauthorblockA{
\textit{International Institute of Information and Technology, Hyderabad, India}\\
\{ishita.vohra, meher.shashwat, aryan.sakaria, amey.kudari\}@alumni.iiit.ac.in\\ nimmi.rangaswamy@iiit.ac.in}
}

\maketitle
\IEEEoverridecommandlockouts
\IEEEpubid{\parbox{\columnwidth}{\vspace{8pt}
\makebox[\columnwidth][t]{IEEE/ACM ASONAM 2022, November 10-13, 2022}
\makebox[\columnwidth][t]{978-1-6654-5661-6/22/\$31.00~\copyright\space2022 IEEE} \hfill}
\hspace{\columnsep}\makebox[\columnwidth]{}}
\IEEEpubidadjcol


\begin{abstract}
The COVID-19 pandemic required eﬀicient allocation of public resources and transforming existing ways of societal functions. To manage any crisis, governments and public health researchers exploit the information available to them in order to make informed decisions, also defined as situational awareness.
Gathering situational awareness using social media, has been functional to manage epidemics. Previous research focused on using discussions during periods of epidemic crises on social media platforms like Twitter, Reddit, or Facebook and developing NLP techniques to filter out important/relevant discussions from a huge corpus of messages and posts. Social media usage varies with internet penetration and other socio-economic factors, which might induce disparity in analyzing discussions across different geographies. However, print media is a ubiquitous information source, irrespective of geography. Further, topics discussed in news articles are already ‘newsworthy’, while on social media ’newsworthiness’ is a product of techno-social processes. Developing this fundamental difference, we study Twitter data during the second wave in India focused on six high-population cities with varied macro-economic factors. Through a mixture of qualitative and quantitative methods, we further analyze two Indian newspapers during the same period and compare topics from both Twitter and the newspapers to evaluate situational awareness around the second phase of COVID on each of these platforms. We conclude that factors like internet penetration and GDP in a specific city influence the discourse surrounding situational updates on social media. Thus, augmenting information from newspapers to information extracted from social media would provide a more comprehensive perspective in resource-deficit cities.
\end{abstract}


\input{introduction}
\input{litReview}

\input{methods}
\input{findings}

\input{discussion}

\scriptsize{

}

\end{document}

%% file: introduction.tex
\section{Introduction}
We explore the application  of situational awareness  \cite{five} as a concept in medical emergencies like the COVID pandemic,  combining knowledge of the dynamic present with pre-existing wisdom to determine likely events that can occur in the near future. Having a perception of the immediate environment and the information around it leads to levels of situational awareness that help people and organizations to respond effectively. Studies have harnessed social media to generate situational awareness in times of need.\looseness=-1
\\
During the second wave of the COVID-19 pandemic in India, people took to Twitter for assistance, especially crowd sourcing medical resources such as oxygen cylinders, hospital beds, and other medical requirements. Despite Twitter not being the go-to social media platform for Indians, the number of tweets from India rose to about 7 times during the peak of the pandemic, in April-May of 2021 as compared to February-March\cite{twenty}. Users  took to Twitter to broadcast  medical crises in their localities, information relating to medical assistance,  and logistical issues that the second wave was causing. Information of this kind  has been analyzed in the past to create situational awareness for specific events and catastrophes. These studies allegedly assume  Twitter to be a comprehensive source of harnessing public discourse around   important topics discussed on any subject in real-time.\cite{twentyfive,twentysix,twentyseven}. This reasoning, in our opinion, maybe acceptable for geographies where the majority of  people use social media as a primary  of communication. However, cities with  limited access to the internet have irregular access to real time information.   Consequently, in  internet constrained areas,   sources such as  print media could likely offer a more comprehensive  public discussion. Prior work has suggested print media as one of the key avenues for taking information to the masses about the spread of the pandemic, institutional curtailment efforts  among other critical issues.\cite{fourtyfour}.
\looseness=-1  
\\
Our study focuses on the following: 
\begin{itemize}
\item It investigates data-augmentation over and above the twitter platform from print media sources  in creating situational awareness of an event 
\item We specifically look into social media and print media and their combined augmentation during the second wave of the COVID crisis. To our knowledge, there is no prior research combining the two media formats to this end. 
\item We explore  Twitter data and the Times of India (the largest English print newspaper in India) and Hindustan Dainik (a newspaper in the Hindi Language) as our news sources, to understand  public discourse surrounding situational updates.
\item Using the data collected,  we compare data augmentation in multiple cities with differing socio-economic factors. 
\end{itemize}
\looseness=-1
To summarize our main contribution is as follows: We use data from Twitter and  traditional print news media, to understand the diversity of information that each of them offered during the second wave of COVID pandemic using  NLP and manual techniques.  Further, we augment the data across the two  media for six major Indian cities. These cities are diverse socio-demographically and in their usage of both  print media and Twitter. We further check if  situational updates and crisis information from Twitter is dependent on the city users tweet from and if the corresponding  use of  augmenting print media indeed offers new data not found on Twitter.  Our study  adds nuance by linking the characteristics of a city, factors like GDP and literacy rate, and the bearing these have on the discussion topics related to the second COVID wave on Twitter  in that city. Thus, our paper not only undertakes a comparative perspective on Twitter and traditional media but augments the study through a second order analysis of the characteristics of a city and Twitter use.
\looseness=-1 

%% file: litReview.tex
\vspace{-5pt}
\section{Literature Review}
This section will briefly discuss literature that dealt with the interplay of COVID-19 and the  information it engendered on social media platforms. During the lockdown in India, social media usage had increased by record numbers,  \cite{fourtynine} subsequently aiding researchers to tap into and analyse a vast pool of data. In this section we  look at past studies using social media platforms to create situational awareness.\looseness=-1

\subsection{COVID-19 and social media platforms}
The sheer amount of online activity  generated due to the COVID-19 pandemic has 
prompted researchers to publish large-scale data sets across various domains  \cite{eleven}. More specific data sets relating to the pandemic have been released. Datasets of tweets specifically on vaccine hesitancy\cite{four}, panacea dataset containing COVID'19 tweets from all over the world \cite{three} and a  multilingual twitter data set containing tweets in different languages from India \cite{twentyone} are studies that served to collect and filter data    for our study.  A study of sentiment analysis of citizens via Twitter data\cite{twentytwo} has limited their study to India,  while  a study on public sentiment and opinions on tweets were collected globally\cite{eight}. The latter  also studied sentiment change over time. Twitter usage were studied during government imposed lockdowns\cite{two} and geospatial trends with regards to sharing conspiracy theories and misinformation \cite{nineteen}. These studies focused on key points useful to our own  and helped delve deeper into our understanding of the interplay between time, socio-economic factors, and sentiments.\looseness=-1

\subsection{Situational awareness, social media and media coverage}
The research community has often studied trends on social media to extract situational awareness when it comes to large-scale events such as natural disasters
 \cite{twentyseven,thirtyfive}, mass emergencies \cite{thirtyfour, fourtyfive}, and more recently the COVID-19 pandemic \cite{thirtyone}. Methods to detect signals for H1N1 outbreak using crowd sourced data  \cite{ten}, quantitative correlations with public health data by applying a topic model to health-related tweets\cite{six} were reported. A study undertook  mobility patterns before and after the WHO declared the pandemic using Twitter data,\cite{twentythree} while another examined the communication between government agencies and citizens on Twitter \cite{twentyfour}. Causal links between media coverage and surge in COVID-19 cases were also studied. These studies further investigated how media coverage mitigated the spread of COVID-19 \cite{fourtysix}. Finally, multi-modal fusion based studies that combine various modes of information for situational awareness have  been undertaken \cite{twentysix,twentyseven}. \looseness=-1
\\
Previous studies have investigated  the creation of  situational awareness explicitly using newspapers to confirm  influence of mass media  on public opinions and reactions in the context of H1N1\cite{fifty} and Ebola\cite{fiftyone}. 
However, to the best of our knowledge, no study has comprehensively investigated situational awareness from both Twitter and print media, specifically for the COVID-19 pandemic in India. In this study, we aim to cover this gap.
 \looseness=-1

%% file: methods.tex
\vspace{-5pt}
\section{Methodology}
\subsection{Data Collection}
We analyze and compare the discussions around COVID-19 during the second wave in India on online social media platform Twitter and two offline print media sources from India among the top 10 newspapers by readership, \cite{twentyfour} Times of India \cite{twelve} and Hindustan Dainik.\looseness=-1 \cite{twentynine}

\subsubsection{Time period considered}
In order to consider a representative time period during the second wave and obtain a substantial amount of discussion on the issue, we fit a Gaussian curve on the graph of daily new cases vs. dates where the peak of the wave corresponds to the center of the curve. Following this, we took dates that are within two standard deviations from the center, as the considered time period for our study. The date range considered is from 11th April 2021 to 30th May 2021. This accounts for the buildup to the peak as well as the decline from the peak.\looseness=-1

\subsubsection{Twitter}
We consider English language tweets from India during the mentioned time period for our study. For collecting tweets, we first obtained unique tweet IDs for all COVID-19-related tweets on Twitter from the Panacea dataset \cite{three} and  processed them using the Twarc API.
\footnote{\url{https://twarc-project.readthedocs.io/en/latest}} API. To get English tweets from accounts in India, we utilized the location field present in the tweet metadata. Extracting via geo-coordinates is a challenging task since it is an optional feature for users and less than 0.5\% turn on their location services due to privacy concerns \cite{fiftytwo}. 
Hence, we collected a mapping of Indian cities to states and then did a fuzzy-wuzzy \footnote{\url{https://pypi.org/project/fuzzywuzzy/}} string comparison of the location field with this mapping. Thus for every Indian tweet, we obtained the corresponding geographic location and city. \looseness=-1

\subsubsection{Newspapers}
To collect news articles we used the online archives of the Times of India and Hindustan Dainik. We used the beautiful soup library \footnote{\url{https://beautiful-soup-4.readthedocs.io/}} to scrape news from both the archives and identified location. We further filtered  news articles obtained using keywords for COVID-19 relevant news articles. Prior studies related to extracting situational awareness or crisis communications use keywords like “\#ebola" for Ebola virus \cite{fourtyfive} and "zika" for Zika virus \cite{thirtyfour}. 
Further, to include data regarding the crisis of medical resource unavailability \cite{fourtyeight} at the time, we included keywords relating to Covid treatments. Hence our English keywords for filtering were - ‘corona’, ’covid’, ’covid-19’,’coronavirus’ ,’vaccine’, ’vaccination’, ’covishield’, ’covaxin’, ’oxygen’, ’plasma’, ’cylinder’, ’remdesivir’, ’ventilator’, ’bed’, ’cremation’, ’crematorium’, ’rt-pcr’, ’rtpcr’ For Hindi, we considered the corresponding Hindi translation of the above phrases.\looseness=-1

For our investigation, we used cities that either had the highest number of English news articles or the highest number of tweets, or both. We selected 2 cities in each category. We also ensured that no two cities were from the same state or proximate geographic locations, as they would face similar issues governed by the same  state government administration. Table \ref{table:cities} mentions the cities we considered along with the number of articles and tweets in each city. However, while considering Hindi Newspapers, we only found articles for Delhi, Mumbai and Patna  as the local language newspapers  for other 3 cities in our study are not Hindi.  In the following section, we provide a qualitative and quantitative summary of the data collected and analyzed for the study. The goal is to understand the diversity of information present, the diversity of topics and to investigate whether augmenting differing sources yields an increase or decrease in diversity of information. The quantitative and qualitative measures of information diversity are explained in later sections of the paper. Analyzing which source of media provides the most comprehensive information is beyond the scope of this study.\looseness=-1

\begin{table}[h]
\vspace{-10pt}
\begin{center}
   \caption{Number of news articles and tweets for each city}
   \label{table:cities}
 \begin{tabular}{|c|c|c|c|}
 \hline
\textbf{City} & \textbf{English News} & \textbf{Hindi News}& \textbf{ Tweets} \\ 
\hline
    Delhi     & 201  & 578 & 148768  \\ \hline
    Mumbai    & 192  & 122 & 105133  \\  \hline
    Hyderabad & 187 &  0 & 43459   \\ \hline
    Kokata    & 132 & 0  & 19520     \\ \hline
    Ahmedabad & 201 & 0  & 7530   \\ \hline
    Patna     & 254  & 352 & 5345    \\ \hline
\end{tabular}
\vspace{-15pt}
\end{center}
\end{table}

\subsection{Data Analysis}
We employed two different modes of analytics research for the huge volume of tweets [over 1 million] and the relatively low number of news articles [200 in number]. We used the open coding-grounded theory approach to extract topics from news. Prior work \cite{fourtyfive, thirtyone} suggests that during epidemics, information found on Twitter is either describing symptoms, discussions about prevention or transmission, or treatment and death reports. Such information is useful for people looking for preventive measures, those  already  ill, government and health monitoring agencies. Hence we did a similar analysis of Tweets and News content. We analyzed news articles of 6 cities and extracted sub-topics based on the following:\begin{itemize}
    \item “What is being talked about' - This helps in identifying issues being  debated. \looseness=-1
    \item “Who are the actors involved” - Here we identify key actors driving debates around issues. We found that the second wave of COVID,  mostly involved the public , government bodies, hospitals, judicial bodies like the high court and Supreme court, and private corporations. \looseness=-1
\end{itemize}
\looseness=-1
For extracting topics out of tweets, we first extracted sentence embeddings using the SBERT architecture \cite{sixteen}. We then used KMeans clustering, based on the elbow method, to cluster tweets for every city. This gave us clusters as mentioned in Table \ref{table:clusters}. To analyze clustering results, we used the following  methods
1) plotting the word cloud for every cluster. 2)  getting the count for every hashtag in a cluster and choosing the top 10 hashtags and 3) For every cluster {\it{k}}, we took the mean of all the embeddings present in the cluster,  ${\mu}$, and then for every embedding, {\it{i}} present in the cluster, we computed its cosine similarity score with ${\mu}$ denoted as ${s_i}$ and sorted all the ${s_i}$ in decreasing order. We then manually analyzed top 50 sentences to get a complete sense of the topics discussed in each cluster. We refer to these topics as sub-topics. \looseness=-1 

After extracting the sub-topics out of both news articles and tweets, we created categories to group similar sub-topics into one category. For example, we saw different clusters and news themes for government imposing lockdown, them starting vaccination for 18+ or starting Oxygen Express to deliver oxygen in the critical states of India, all these sub-topics can be clubbed into one single category of “government dealing with COVID’19 crisis". We then analyzed every single category \& sub-topics present in them, for both news and Twitter to find out the differences. Additionally, we cross-checked  sub-topics found on Twitter that don’t occur in our news articles in  TOI/Hindustan Dainik published during the time period. Similarly, for the sub-topics found in print news but not on Twitter we undertook a keyword search of  tweets to understand the rate and intensity of discussion among the identified   sub-topics. Sub-topics involving black-marketing of medicines, suicides committed by COVID patients or cab drivers, and food delivery guys getting infected were  not heavily discussed in some cities on Twitter.\looseness=-1  

\begin{table}
\vspace{-15pt}

\begin{center}
   \caption{Number of clusters generated for each city}
   \label{table:clusters}
  \begin{tabular}{|c|c|}
  \hline
\textbf{City} & \textbf{\#Clusters for Articles}  \\ \hline
    Delhi     & 140       \\ \hline
    Mumbai    & 100    \\ \hline 
    Hyderabad & 75      \\ \hline
    Kolkata    & 30        \\ \hline
    Ahmedabad & 25      \\ \hline
    Patna     & 18       \\ \hline
\end{tabular}
\vspace{-20pt}
\end{center}
\end{table}

\subsection{Quantitative Analysis}
As situational updates on Twitter for a particular city will depend on the number of people using the platform in that city, we decided to investigate the co-relation in the diversity of information extracted from news articles and Twitter data and a further correlation with the socio-economic characteristics of a city. We do a quantitative study as follows: For each city, we obtain the set of all unique keywords discussed on the English news (\textit{\textbf{N}}) and English twitter (\textit{\textbf{T}}), along with the frequency of appearance for each keyword. We pre-processed both news and tweets. For both news articles and tweets, we removed stop words, non-ASCII characters, punctuation followed by lemmatization and lower casing. For tweets, we removed the links, user mentions and hashtags. Considering the high number of Twitter keywords and to filter out scattered one-off mentions, we consider tweets with keywords that have a minimum frequency of 10.\looseness=-1

In order to study this quantitatively, we calculate two metrics -
\begin{itemize}
\item {\verb|Metric 1|}: The Jaccard index (intersection over union) gives us an estimate of the overlap between news and twitter keywords. 
\begin{displaymath}
 \frac{T \cap N}{T \cup N}
\end{displaymath}

\item {\verb|Metric 2|}: We also define a metric to quantify news keywords left out from twitter discussions. This is the ratio of unique news topics (only discussed on the news) to all the news topics. A lower value would enable Twitter data to bring out more topics discussed on the news and vice versa
\begin{displaymath}
 \frac{N - (T \cap N)}{N}
\end{displaymath}

\end{itemize}

\subsection{City Profiling}
For cities, we consider socio-economic attributes like GDP (in \$Bn), internet penetration (IP in \%) , literacy (Lit. in \%), population (in millions) and the  number of COVID cases as shown in Table III. We investigate if the disparity in situational updates on Twitter and print media could be attributed, even if partially, to the location we are studying. \looseness=-1
We consider two cities with the highest GDP in India i.e. Delhi and Mumbai. These coincidentally have the highest number of Tweets (as seen in Table \ref{table:cities}), which can be attributed to higher internet penetration. Moreover, Mumbai is the financial and business capital of the country, while Delhi is the political capital and home to the central government. These cities are also India’s largest metropolitan areas \cite{twentyeight} bearing the highest number of COVID cases in India (Delhi - 711,169) and (Mumbai - 194,110). Ahmedabad and Patna saw fewer tweets overall but a significant number of news articles. Ahmedabad has a much higher GDP than Patna, the capital of one of India’s least developed states, Bihar \footnote{\url{https://en.wikipedia.org/wiki/List_of_Indian_states_and_union_territories_by_Human_Development_Index}} \&
city producing  the least amount of Tweets but a no dearth of news articles both in Hindi and English. This city also has the lowest GDP and internet penetration rate, adding to the diversity of cities we consider for our study. Both Hyderabad and Kolkata have a comparable amount of Tweets and News Articles. From Table III, it can be shown that Kolkata has double the population, higher literacy and significantly higher GDP. In the next section, we explore city wise situational updates, government  action and critical information circulation. \looseness=-1

\begin{table}
\vspace{-15pt}
   \label{table:profile}
\caption{Socio economic factors of cities considered. GDP refers to Gross Domestic Product in \$B, IP refers to Internet Penetration in \%, Lit refers to Literacy in \%, Populn refers to Population in \%.}
\begin{tabular}{|c|c|c|c|c|c|}
\hline
\textbf{City} & \textbf{GDP} & \textbf{IP} & \textbf{Lit.} & \textbf{Populn.} & \textbf{\#Cases}\\ \hline
Delhi & 210 & 61.068 & 88.7 & 25.83 & 711169 \\ \hline
Mumbai & 240 & 55.820 & 84.8 & 25.97 & 194110 \\ \hline
Kolkata  & 150 & 41.216 & 80.5 & 18.54 & 153875 \\ \hline
Hyderabad & 74 & 39.886 & 72.8 & 8.61 & 145324 \\ \hline
Ahmedabad & 80 & 47.270 & 82.4 & 7.78 & 153711 \\ \hline
Patna & 30 & 31.693 & 70.9 & 3.60 & 84944 \\ \hline
\end{tabular}
\label{tab1}
\vspace{-15pt}
\end{table}

%% file: findings.tex
\vspace{-5pt}
\section{Findings}
As discussed in our research agenda, we found disparity in   topics discussed in the news and Twitter. We found  eight major categories of missing topics which include government organizations, MNCs/NGOs, Supreme Courts taking actions to deal with the pandemic/crisis, hospitals giving information related to preventive measures, effects on the economy due to lockdown, black marketing, inadequate distribution of hospital resources and doctors involved in malpractices. We further club them into three broad themes which are discussed as follows. \looseness=-1

\subsection{Stakeholders}
In this section, we will briefly discuss the response to the pandemic by specific stakeholders of the Indian socio-economic system and the discourse surrounding them. We will limit our discussion to the actions of the central and state governments, the Supreme Court, Doctors, Police, and certain organizations. We have filtered these specific stakeholders, as the majority of the discourse found on both print media and Twitter was centered around them. The actions taken by these stakeholders are critical  for people affected by the pandemic, hence, actions and decisions made by key stakeholders can make a profound difference in the affected populations. In the following subsections, we will briefly discuss the actions and decisions of these stakeholders in the cities under study. We will further correlate the discourse generated as a direct result of the actions of the stakeholders and point out the differences in information found in our two comparative mediums: print and online. Additionally, we will report certain socio-economic markers(GDP, Internet penetration, etc) that can provide more context around both  decisions made by the specific stakeholders as well as the topics from the emergent discourse. \looseness=-1

\subsubsection{Government's handling of the crisis}
The pandemic was the first of its kind, given the scale at which disrupted the everyday of society. Governments at all levels were expectedly unprepared to deal with the crisis. Central and state governments provided aid to citizens in myriad ways, such as setting up oxygen cylinder plants, launching community kitchens, and setting up medical teams to name a few. They also provided family pensions and government jobs to family members of doctors and health personnel who succumbed to death. While help was provided in all the cities, discourse for the same showed a varying disparity between Twitter and print media among the cities considered. In the city of Patna, the capital of Bihar, tweets discussing the aid provided by the local and state government were low in number. This is potentially because internet penetration is lower in Patna compared to the other cities, and consequently lower number of Twitter users. In contrast, the aforementioned aid provided in Patna was covered widely in print media (newspapers). Cities like Mumbai and Delhi (both have high GDP) show similar discussions on Twitter and English print media. However, Hindi news was able to bring out topics not mentioned in both. For example, Hindi news covered arrangements made for special trains to transport migrant workers back to their hometowns. Hindi news media also discussed the testing facilities setup for early detection that were instrumental for curbing the initial rise in cases. Localised rules and reforms were rolled out in cities to manage any crisis specific to those cities. For instance, in Kolkata, cases rose concurrently with election time and  rules regarding overcrowding at political rallies were put in place. Discourse around the above  was low on Twitter, but covered by the print media. In some cases, we also notice a difference in sentiments towards the government. For example - Tweets in Mumbai reflect the oxygen shortage and people complaining about the same. However, print media (especially Hindi news) reflects positively on the local government appreciating the Mumbai Municipal Corporation. Oﬀicials were praised for using oxygen tanks  installed in the first wave. \looseness=-1

\subsubsection{Doctor's handling of the crisis}
Twitter was less widely used in cities like Ahmedabad and Patna as was reflected in the count of total tweets sampled from these regions. As a result, medical advice was primarily conveyed through print media in these cities. Medical professionals used media outlets to spread awareness regarding complications such as black fungus and methods to treat them amongst medical staff that were operating in rural areas. Even in a city like Mumbai, which has had a high overlap in information found on Twitter and in print media, there was information regarding medical treatments in print media  not very prevalent in tweets. For example, information regarding Remdesivir tablets and its efficacy on other factors such as oxygen therapy, vitamin supplements, steroids  were found in news articles, in contrast to  the  4 tweets out of hundreds of thousands in the Mumbai region. From the news articles, it appears  youth and women were ignored while providing COVID care in cities like Delhi (the city which produced the maximum cases in India) whereas Twitter does not mention any such scenario. Help was given by hospitals to these sets of people by launching specialized care units. We see a similar scenario in Patna, where the second wave was harsh for children and hence doctors took special measures directed at the Youth by increasing beds and ICU units in the pediatric wards. Such type of information is missing from Twitter, thus excluding an entire demographic. \looseness=-1

\subsubsection{High Court/Supreme Court handling of the crisis}
 In almost all cities the Supreme Court and High Court intervened to take matters into their hands and make decisions as the government faltered in handling the crisis. In acute contexts of stress like Delhi, there are news articles about high court directing all retailers to cap the price of medical equipment below   retail prices and also  suggested the Delhi Government provide oximeters for economically weaker sections. However there is no discussion for these topics on Twitter. A team of medical oﬀicers were summoned   to perform audits for the supply and distribution of oxygen in various zones. This sub-topic has 30/148768 tweets talking about the audits. Newspapers draw attention to irregular practices across cities during the second wave:  in Hyderabad, police oﬀicials stopped  ambulances coming outside Telangana district, to avoid increasing cases in the region. In Kolkata, the high court strictly enforced rule flouting norms  during poll campaigns. However, there is no discussion on Twitter on the above. In Ahmedabad, the high court insisted that the state government give priority to vaccinating those wait listed for the second dose of vaccine and questioned the availability of doses, resource shortage, planning , management and distribution strategies. \looseness=-1

\subsubsection{MNCs/NGOs handling of the crisis}
Print media mentions that  the city of Hyderabad, covered telemedicine services were launched by the MNCs/NGOs as an outreach strategy. Reports also mention the work of social activists extending help to book slots for vaccine doses and helping people register on the official CoWin platform.\footnote{\url{\https://www.cowin.gov.in/}}
(India's vaccine registration portal) Reports mention the police launching aid programs, websites for plasma donation and oxygen tanks to help the resource poor. All of the above topics do not find a mention on Twitter. \looseness=-1

\subsection{Resource Management}
Here we discuss data around management of medical resources like oxygen cylinders, beds and Remdesivir injections throughout the pandemic. From both News and Twitter we observe with the rise in cases, there was a decline in medical supplies, fostering an environment for black marketing and doctors engaging in malpractices selling medical supplies at high prices and admitting fake patients. In the following subsections we detail how discussions around resource management vary between cities and  the two platforms. \looseness=-1

\subsubsection{Repletion of hospital resources}
We see cities like Ahmedabad and Patna, publicizing a large number of news articles capturing the “why” and “how” of  resource mismanagement and an absence of mention on  Twitter. Ahmedabad reported the Remdesivir crisis to the improper distribution of injections amongst patients. Reports touched upon medical care extended to non-critical patients leading to a shortage of supply amongst critical patients. COVID beds and oxygen supplies were extended to asymptomatic patients. Doctors went on strike due to the government not responding to their demands of increasing wages thereby causing a shortage of treatment. None of these saw mentions on the Twitter platform. \looseness=-1

\subsubsection{Malpractices by Doctors}
This particular sub-theme is only discussed in the print media for all cities with no mention  on Twitter. In Kolkata, hospitals created an artificial shortage of medical supplies forcing the public  to buy at exorbitant prices charged by local pharmacists. Ahmedabad and Patna were reported charging high rates for hospital beds, oxygen cylinders beyond the government cap. Delhi hospitals indulged in  communicating false information about the availability of beds and selling drugs at astronomical prices. \looseness=-1

\subsubsection{Black Marketing} 
In high GDP cities like Mumbai and Delhi, there is an active discussion concerning black marketing of oxygen cylinders, Remdesivir vials, and COVID medicines being sold at exorbitant prices on both Twitter and Newspapers. However, in cities like Ahmedabad (GDP being comparatively lower) and Patna (with a much lower GDP), we see little to no discussion (Ahmedabad only 2 out of 7530 tweets reflect this and Patna showed zero tweets). A specific case in Kolkata, oxygen cylinders were sold at exorbitant prices and arrests were made- interestingly there is a  Twitter discussion involving the black-marketing, but none on the arrests. \looseness=-1

\subsection{ Economic Impact of COVID}
In this section, we analyze discussions around the effects that the pandemic had on the economy. We undertake a comparison of discussion topics on twitter and print media,  the location of this data contributing towards disparity in discussions on both mediums. In high GDP cities like Delhi and Mumbai, we gather a similar picture about the impact on businesses and economic losses incurred, on both News and Twitter. However, in lower GDP cities like Patna and cities having a lower number of Tweets like Ahmedabad, we notice a disparity in discussions about the same issue between News and Twitter. Discussions around the government-imposed restrictions on the number of guests allowed at weddings and social gatherings in Patna and the ensuing losses suffered by the  food and beverage industry were reported only in the print news media.  In Kolkata, newspapers reported the fear disseminating in the airline industry about low flight loads caused by mandatory RT-PCR testing for air travellers.Twitter offered no discussion on topics related to the airline industry. \looseness=-1

\subsection{Quantitative Analysis}
Differences in situational updates across cities reported in the print media news and on	social media can be attributed to key socio-economic factors among these cities.  Table \ref{table:metrics} explains the two metrics used to quantify the coverage of news topics in Twitter discussions and estimating the overlap between situational updates discussed on Twitter and news for every city
\looseness=-1
\begin{table}[h]
\vspace{-15pt}
\begin{center}
   \caption{Metrics computed}
   \label{table:metrics}
  \begin{tabular}{|c|c|c|}
  \hline
\textbf{City} & \textbf{Metric 1} & \textbf{Metric 2}\\
   \hline
    Delhi     & 0.328    & 0.474    \\ \hline
    Mumbai    & 0.325    & 0.488       \\  \hline
    Hyderabad & 0.322    & 0.604   \\ \hline
    Kokata    & 0.250    & 0.717       \\ \hline
    Ahmedabad & 0.147    & 0.849   \\ \hline
    Patna     & 0.080    & 0.919  \\ \hline
\end{tabular}
\vspace{-7pt}
\end{center}
\end{table}

We computed the cross-correlation of each metric with factors like GDP and internet penetration to explore socio-economic factors influencing disparity in discussion of keywords. From Table \ref{table:corre}, we noted strong correlations between our metrics and
socio-economic factors. 

\textbf{GDP}: There is a high positive correlation between GDP and Metric 1. This signifies that there is a high overlap of keywords discussed on twitter an in the news in areas with high GDP. There is also a strong negative correlation with Metric 2, which denotes very few news keywords that don’t reach twitter discussions. Both these metrics denote that twitter can be used as a reliable source for analyzing discussions in cities with higher economic prosperity, simultaneously displaying  very low values for Metric 1 and a high value for Metric 2 (from Table  \ref{table:corre} in cities with low GDP, like Patna).\looseness=-1

\textbf{Internet Penetration}:  There is a high positive correlation with Metric 1, which can be attributed to more number of people having access to the internet. There is also a strong negative correlation with Metric 2, which denotes that there are very few news keywords that don’t reach twitter discussions. Both these metrics denote that twitter can be used as a reliable source for analyzing discussions in areas with high internet penetration. Places with low internet penetration, for example Patna, reveal disparity in discussions on Twitter and the news. In such places including offline platforms like news in the study becomes imperative to fully capture public discussion. \looseness=-1

\begin{table}[h]
\vspace{-10pt}
\begin{center}
   \caption{Correlation of Metrics with Socio-Economic Factors}
   \label{table:corre}
  \begin{tabular}{|c|c|c|}
  \hline
\textbf{City} & \textbf{Metric 1} & \textbf{Metric 2} \\
   \hline
    Gross Domestic Product     & 0.743    & -0.842   \\ \hline
    Internet Penetration      & 0.659    & -0.785   \\ \hline
\end{tabular}
\end{center}
\vspace{-10pt}
\end{table}
Adopting qualitative and quantitative analysis helped us examine not only the differences in the nature of information but also illumining  the disparities in the situational information discussed in different cities over the two kinds  of media. Based on the type of city, the benefits of augmenting news and social media vary. In cities like Delhi and Mumbai, where people use the Twitter platform most  Twitter discussions overlap with what is present in the news. However, in cities like Patna or Ahmedabad with weak Twitter use, augmenting both types of media brings diversity to media information and situational awareness.\looseness=-1

%% file: discussion.tex
\vspace{-5pt}
\section{Discussion}
From our findings we observe disparity in topics discussed on Twitter and News across different cities. For example, certain topics discussed in the news are either completely absent or hardly discussed on Twitter. We further see a strong correlation between the above disparity in the news and on Twitter and socio-economic factors [like GDP and Internet Penetration] in the cities that are reporting news or tweeting. We found a high positive correlation between GDP of a city and the overlap of discussion on Twitter and News. There are several topics discussed significantly on both platforms. These range from governments imposing lock downs, hospitals setting up COVID care centers, state governments coming forward and providing oxygen tanks, reporting the number of new cases/deaths and positive rates, vaccine shortage, encouragements to get vaccinated, and political parties blaming each other for rising cases (However, the last topic is significant on Twitter than in print news). There is an adequate amount of news offering information that is absent from Twitter. In resource-poor places, where internet penetration is low, we observe that news is more representative of the lives of common people. The news is more localized that Twitter fails to represent, especially in cities with weaker internet penetration and lower GDP.\looseness=-1
\\
A striking similarity which we observed in  high-income (Delhi) and low-income (Patna) city revolves around doctors targeting resources mostly towards affected children  during the second wave. We don’t see tweets around this occurrence. Further, we see news articles cover communities who are otherwise missing on Twitter. For example, several news articles report the status of blue-collar workers, the socially and marginalised like Maoists] during the pandemic while Twitter has no mention of them. News articles report  NGOs providing help to remote areas and launching special services that aided vaccinations. Moreover, many doctors gave treatment-related advice regarding the COVID pandemic in newspapers rather than on Twitter. Malpractices by hospitals creating artificial shortages in oxygen cylinders and selling at higher prices are some of the topics mentioned only in print media irrespective of the city. 
\\
Further, we had observed the cities of Patna, and Ahmedabad, overlap in the number of tweets about the Delta wave and share similar discussion topics in news articles and the Twitter platform. For instance, doctors offered COVID-related advice in the news instead of tweeting about it, and news articles captured how and why medical supplies get replenished and absent on the Twitter Platform. The latter broadly reflected shortage of resources. The above cities are otherwise socio-economically varied  and our findings indeed suggest that discourse is varied in a few cases. For example - news articles bring out the state government focusing on inoculation of disabled people in Ahmedabad. We don’t see such articles in Patna. Our findings suggest that for some topics, the extent of overlap of discussions in digital and print media indeed depends on which city is being considered. We see discussions around the lack of oxygen cylinders and Remdesivir injections as well as a few other illicit practices covered in print media. However, we observed engaged discussions only in the bigger metropolitan cities of Delhi, Mumbai, and Hyderabad on Twitter. In  metros like Kolkata and Ahmadabad, the intensity of similar discussion is not present (<0.01\% of total tweets) while in Patna, a relatively smaller city, we don't see the above topics being tweeted. These differences can be attributed to the economic development of the cities and varying internet penetration. Further, we observed both English and Hindi news sources have significant overlap in discussion topics, but Hindi news, being more localized, is able to engage with topics that did not surface in the English news. For example, there were incidents of commotion created at hospitals and community kitchens by people due to lack of medical resources and food. Another example highlighted in the Hindi news is about doctors and front-line workers being underpaid, especially those who worked tirelessly during the second wave to treat COVID patients.
\looseness=-1

\vspace{-5pt}
\section{Conclusion}
In this study, we examine two types of media,  print newspapers and the social media platform, Twitter, covering the second COVID wave in India. We suggest that a combined analysis  of both helps to surface diverse  situational information in the context of the second delta wave. Extracting more information in the domain of situational awareness significantly aids organizations and people at large to make decisions that are in their best interest. Moreover, it also helps the post-disease and pre-disease communities deal with pandemics effectively by consolidating information about treatments, symptoms, and aid among others. Our study revealed in cities where internet penetration is low with restricted access to Twitter, print media can help provide information filling the gap that internet-enabled platforms could provide. We highlight that adding regional newspapers assists in enriching the topics found in print media. Our future work would extend this study to other regional languages like Gujarati, Marathi, and Bengali. We also aim to explore augmenting Twitter and Newspapers with live news channels.
Situational updates given by Twitter are real-time, which is not the case for newspapers and until the time infrastructure improves across cities in India, augmented social and print media  remain a good source of situational information. 
\looseness=-1

%% file: sample-base.bbl
\begin{thebibliography}{00}
\bibitem{two} A. Arora, P. Chakraborty, M.P.S Bhatia, and P. Mittal, ``Role of Emotion in Excessive Use of Twitter During COVID-19 Imposed Lockdown in India,`` Journal of Technology in Behavioral Science 2021.
\bibitem{three}J. M. Banda et al., “A Large-Scale COVID-19 Twitter Chatter Dataset for Open Scientific Research—An International Collaboration,” Epidemiologia, vol. 2, no. 3, pp. 315–324, Aug. 2021, doi: 10.3390/epidemiologia2030024.
\bibitem{four} M. R. DeVerna, F. Pierri, B. Truong, J. Bollenbacher, D. Axelrod, N. Loynes, C. Torres-Lugo, K. Yang, F. Menczer, and J. Bryden, ``CoVaxxy: A Collection of English-Language Twitter Posts About COVID-19 Vaccines,''  ICWSM 2021.
\bibitem{five} M. R. Endsley, ``Toward a Theory of Situation Awareness in Dynamic Systems,'' Human Factors Volume: 37 issue: 1, 1995.
\bibitem{six} X. Huang, Z. Li, Y. Jiang, X. Li, and D. Porter, ``Twitter reveals human mobility dynamics during the COVID-19 pandemic,'' PLOS ONE 2020.
\bibitem{eight} J. S.-L. Kwan and K. H. Lim, “Understanding Public Sentiments, Opinions and Topics about COVID-19 using Twitter,” CoRR, vol. abs/2012.03039, 2020, [Online]. Available: https://arxiv.org/abs/2012.03039 
\bibitem{ten} R. Munro, L. Gunasekara, S. Nevins, L. Polepeddi, and E. Rosen, “Tracking epidemics with natural language processing and crowdsourcing,” AAAI Spring Symposium - Technical Report, pp. 52–58, Jan. 2012.
\bibitem{eleven}S. S. Naseem, D. Kumar, M. S. Parsa, and L. Golab, “Text Mining of COVID-19 Discussions on Reddit,” in 2020 IEEE/WIC/ACM International Joint Conference on Web Intelligence and Intelligent Agent Technology (WI-IAT), 2020, pp. 687–691. doi: 10.1109/WIIAT50758.2020.00104.
\bibitem{twelve}T. of India(TOI), “Time of India article archive,” 2022, [Online]. Available: https://timesofindia.indiatimes.com/archive.cms
\bibitem{sixteen}N. Reimers and I. Gurevych, “Sentence-BERT: Sentence Embeddings using Siamese BERT-Networks,” CoRR, vol. abs/1908.10084, 2019, [Online]. Available: http://arxiv.org/abs/1908.10084
\bibitem{nineteen}M. Stephens, “A geospatial infodemic: Mapping Twitter conspiracy theories of COVID-19,” Dialogues in Human Geography, 2020, doi: 10.1177/2043820620935683.
\bibitem{twenty}H. Times, “Tweets soared 7 times higher during Covid-19 second wave peak,” Jun. 2021, [Online]. Available: https://www.hindustantimes.com/india-news/tweets-soared-7-times-higher-during-covid-19-second-wave-peak-twitter-101625072665359.html
\bibitem{twentyone} Uniyal, D., \& Agarwal, A. (2021). IRLCov19: A Large COVID-19 Multilingual Twitter Dataset of Indian Regional Languages. CoRR, abs/2107.12360. https://arxiv.org/abs/2107.12360 
\bibitem{twentytwo} A. S. M. Venigalla, D. Vagavolu, and S. Chimalakonda, “Mood of India During Covid-19 - An Interactive Web Portal Based on Emotion Analysis of Twitter Data” https://arxiv.org/abs/2005.02955
\bibitem{twentythree}{Y. Wang, H. Hao, and L. S. Platt, “Examining risk and crisis communications of government agencies and stakeholders during early-stages of COVID-19 on Twitter” doi: https://doi.org/10.1016/j.chb.2020.106568.}
\bibitem{twentyfour} A. Bureau, “Newspaper Count,” Feb. 2022. http://www.auditbureau.org/files/JJ2018
\bibitem{twentyfive} M. Dalili Shoaei and M. Dastani, "The Role of Twitter During the COVID-19 Crisis: A Systematic Literature Review," Acta Informatica Pragensia, vol. 9, pp. 154-69, 2020.
\bibitem{twentysix}X. Gui, Y. Kou, K. H. Pine, and Y. Chen, “Managing uncertainty,” May 2017.
\bibitem{twentyseven} S. Vieweg, A. L. Hughes, K. Starbird, and L. Palen, “Microblogging during two natural hazards events,” 2010.
\bibitem{twentyeight} City Mayors: World’s Largest Urban Areas in 2020. \texttt{\url{http://www.citymayors.com/statistics/urban_2020_1.html}}\looseness=-1 
\bibitem{twentynine}“Live Hindustan,” https://www.livehindustan.com/.
\bibitem{thirtyone}Y. Wang, H. Hao, and L. S. Platt, “Examining risk and crisis communications of government agencies and stakeholders during early-stages of COVID-19 on Twitter,” Comput. Human Behav., vol. 114, no. 106568, p. 106568, Jan. 2021.

\bibitem{thirtyfour}L. Hagen, S. Neely, R. Scharf, and T. E. Keller, “Social media use for crisis and emergency risk communications during the Zika health crisis,” Digit. Gov.: Res. Pract., vol. 1, no. 2, pp. 1–21, Apr. 2020.
\bibitem{thirtyfive}A. Karami, V. Shah, R. Vaezi, and A. Bansal, “Twitter speaks: A case of national disaster situational awareness,” J. Inf. Sci., vol. 46, no. 3, pp. 313–324, Jun. 2020.
\bibitem{fourtyfour}G. A. Parvin, R. Ahsan, M. H. Rahman, and M. A. Abedin, “Novel Coronavirus (COVID-19) pandemic: The role of printing media in Asian countries,” Front. Commun., vol. 5, Nov. 2020.
\bibitem{fourtyfive} K. Rudra, A. Sharma, N. Ganguly, and M. Imran, “Classifying Information from Microblogs during Epidemics,” Jul. 2017. [2]H. Zade, K. Shah, V. Rangarajan, P. Kshirsagar, M. Imran, and K. Starbird, “From situational awareness to actionability,” Proc. ACM Hum. Comput. Interact., vol. 2, no. CSCW, pp. 1–18, Nov. 2018.
\bibitem{fourtysix} Weike Zhou,  Aili Wang,  Fan Xia,  Yanni Xiao,  Sanyi Tang. Effects of media reporting on mitigating spread of COVID-19 in the early phase of the outbreak[J]. Mathematical Biosciences and Engineering, 2020, 17(3): 2693-2707. doi: 10.3934/mbe.2020147
\bibitem{fourtyeight}
"Lack of planning led to shortage of drugs, equipment during COVID second wave in Kerala", 2022. [Online]. Available: https://www.thehindu.com/news/national/kerala/lack-of-planning-led-to-shortage-of-drugs-equipment-during-covid-second-wave-in-kerala-say-health-officials/article34823066.ece.
\bibitem{fourtynine} Twittersees-record-number-of-users-during-pandemic-but-advertising-sales-slow", Washington Post, 2022. [Online]. Available: http://twitter-sees-record-number-of-users-during-pandemic-but-advertising-sales-slow.
\bibitem{fifty}
Chang C. News coverage of health-related issues and its impacts on perceptions: Taiwan as an example. Health Commun. 2012;27(2):111-123. doi:10.1080/10410236.2011.569004
\bibitem{fiftyone}
Adelakun, Lateef \& Adnan, Hamedi. (2016). Communicating Health: Media Framing of Ebola Outbreak in Nigerian Newspapers. Malaysian Journal of Communication. 32. 10.17576/JKMJC-2016-3202-19. \bibitem{fiftytwo}Li, Rui, Shengjie Wang, Hongbo Deng, Rui Wang and Kevin Chen-Chuan Chang. “Towards social user profiling: unified and discriminative influence model for inferring home locations.” KDD (2012).
\looseness=-1
\end{thebibliography}
